\documentclass[aps,prc,twocolumn,showpacs,superscriptaddress,groupedaddress]{revtex4}  

\usepackage{amsmath}
\usepackage{amssymb}
\usepackage{amsfonts}
\usepackage{float}
\usepackage{graphicx}
\usepackage{tabularx}
\usepackage{multirow}
\usepackage{lscape}
\usepackage{rotating}
\usepackage{pstricks}

\def\be{\begin{equation}}
\def\ee{\end{equation}}
\def\bea{\begin{eqnarray}}
\def\eea{\end{eqnarray}}
\def\bean{\begin{eqnarray*}}
\def\eean{\end{eqnarray*}}
\newcommand{\bk}{\boldsymbol{\mathbf k}}
\newcommand{\bkp}{\boldsymbol{\mathbf k^\prime}}
\newcommand{\vsigma}{\boldsymbol{\mathbf\sigma}}
\newcommand{\br}{\boldsymbol{\mathbf r}}
\def\nn{\nonumber}
\def\nnn{\nonumber \\}

\begin{document}

\title{Partial wave decomposition of finite-range effective tensor interaction}

\author{D. Davesne}
\affiliation{Universit\'e de Lyon, F-69003 Lyon, France; Universit\'e Lyon 1,
             43 Bd. du 11 Novembre 1918, F-69622 Villeurbanne cedex, France\\
             CNRS-IN2P3, UMR 5822, Institut de Physique Nucl{\'e}aire de Lyon}
\author{P. Becker}
\affiliation{Universit\'e de Lyon, F-69003 Lyon, France; Universit\'e Lyon 1,
             43 Bd. du 11 Novembre 1918, F-69622 Villeurbanne cedex, France\\
             CNRS-IN2P3, UMR 5822, Institut de Physique Nucl{\'e}aire de Lyon}
\author{A. Pastore}
\affiliation{Department of Physics, University of York, Heslington, York, Y010 5DD, UK}
             \author{J. Navarro}
\affiliation{IFIC (CSIC-Universidad de Valencia), Apartado Postal 22085, E-46.071-Valencia, Spain}

\begin{abstract}
We perform a detailed analysis of the properties of the finite-range tensor term associated with the Gogny and M3Y effective interactions. In particular, by using a partial wave decomposition of the equation of state of symmetric nuclear matter, we show how we can extract their tensor parameters directly from microscopic results based on bare nucleon-nucleon  interactions. Furthermore, we show that the zero-range limit of both finite-range interactions has the form of the N3LO Skyrme pseudo-potential, which thus constitutes a reliable approximation in the density range relevant for finite nuclei. Finally, we use Brueckner-Hartree-Fock results to fix the tensor parameters for the three effective interactions. 
\end{abstract}


 
\date{\today}

\maketitle

One of the key ingredients of the bare nucleon-nucleon (NN) interaction is the tensor part, which represents the most distinct manifestation on meson exchange process. Among the most important properties related to the tensor interaction, one can mention the quadrupole moment of the deuteron, the properties of excited states~\cite{bro06,les07,bai09,min13}, the contribution to the spin-orbit splitting~\cite{ari60}, and the shell evolution along isotopic chains~\cite{Ots05}.  Within the context of phenomenological interactions, one should in particular keep in mind that the strong competition between spin-orbit and tensor on the shell structure properties of atomic nuclei~\cite{les07,zal08,zou08,ben09}, prevents to freely explore the parameter space, since a variation in the tensor can lead to a substantial change in the shell structure with the appearance of new magic numbers. A detailed discussion and list of references about these issues is given in the recent review article of Sagawa and Col\'o~\cite{sag14}.

Despite their importance, only a few exploratory attempts have been made in the past~\cite{sta77,oni78}  to include explicit tensor terms within non-relativistic self-consistent mean-field or density functional nuclear models. 
More recently, such terms have been added to existing zero- or finite-range effective interactions, as the popular families of Skyrme~\cite{vau72}, Gogny~\cite{dec80} or M3Y~\cite{ber97}. The parameters are usually fixed to reproduce some selected spin-orbit splittings, with either a partial or a complete fit of parameters. However, the parametrizations can lead to the appearance of unphysical instabilities of the Fermi surface both in the zero- and long-range regimes~\cite{cao10,nav13,pas15}. To avoid them we have proposed~\cite{dav14P,dav15prep} to incorporate into the fitting procedure the constraints obtained from linear response theory. Furthermore, we have also suggested~\cite{pas14L,Dav15,dav14P} some new constraints from microscopic calculations based on the bare NN interaction: specifically, a partial wave decomposition on the so-called N3LO Skyrme pseudo-potential of the symmetric nuclear matter (SNM) equation of state (EoS) allows one to clearly identify the contribution of each term of the effective interaction, even for the tensor part. This decomposition can be used for an initial guess of the interaction parameters. The N3LO pseudo-potential represents the most general non-local zero-range pseudo-potential that includes all possible terms up to the sixth order in derivatives, that is up to the next-to-next-to-next leading order. It has been constructed several years ago~\cite{car08,rai11} and has been recently written explicitly in the more familiar Cartesian basis, constraining it to be gauge invariant~\cite{dav14a,dav14b}. In this approach, the standard Skyrme interaction with tensor terms corresponds to N1LO plus the usual density-dependent term.

In the present article, we use as a guide the partial wave decomposition of the symmetric nuclear matter EoS to isolate the spin-orbit and tensor contributions of some finite-range interactions. Specifically, we consider the finite-range interactions D1ST2a~\cite{gra13} and M3Y-P2~\cite{nak03}, and the N3LO pseudo-potential as well. We show that new constraints are obtained on the sign (even the sign is not known with certainty) and the order of magnitude of the tensor parameters.  

The potential contribution to the SNM energy per particle can be decomposed by using a coupled spin-isospin $(S,T)$ basis as
\begin{equation}
\frac{E}{A}=\frac{3}{5}\frac{\hbar^2 k_F^2}{2m}+\sum_{ST} {\mathcal{V}^{(ST)}}(k_F) \, ,
\end{equation}
where $k_F=(3 \pi^2 \rho/2)^{1/3}$, $\rho$ being the density, and $\mathcal{V}^{(ST)}$ is the potential energy per particle projected onto the different channels.  Neither the tensor nor the spin-orbit interactions contribute to the $(S,T)$ components. To study them one has to go one step back in the calculation and project the potential energy terms onto the $J,L,S,T$ subspaces, where $J$ and $L$ are the total and orbital angular momenta, respectively. Using the standard spectroscopic notation $^{2S+1}L_J$, the EoS is written as
\begin{equation}
\frac{E}{A}=\frac{3}{5}\frac{\hbar^2 k_F^2}{2m}+\sum_{JLS} {\mathcal{V}}(^{2S+1}L_J) \, ,
\end{equation}
where for simplicity the explicit $k_F$-dependence in the potential energy terms has been suppressed. The value of $T$ is immediately deduced from the antisymmetry of the matrix elements. Adding up ${\mathcal{V}}(^{2S+1}L_J)$ for all $J$ and $L$ values, one gets $\mathcal{V}^{(ST)}$, with an exact cancellation of tensor and spin-orbit contributions. 

Our analysis is based on Brueckner-Hartree-Fock (BHF) calculations~\cite{bal97} of the partial wave contributions to the potential energy per particle, derived  from the microscopic Argonne $v14$ nucleon-nucleon two-body interaction plus the Urbana model for the three-body term. Although the SNM saturation point given by this calculation is slightly shifted to higher values of density and energy per particle, these results provide us with a helpful guide to assess the properties of effective interactions in a mean-field scheme. Other microscopic calculations exist, as low-$k$ chiral effective field ($\chi$-EFT)~\cite{heb11} or the many-body perturbation theory (MBPT)~\cite{bog05,rot08}, but all of them have been limited to the $(S,T)$ channels. In this respect it is worth noticing that for these channels all these calculations agree below saturation density. 

To make a proper comparison of the tensor and spin-orbit contents of the effective interactions we realise that 
for a given $L, S$ pair, the central part of the interaction gives the same contribution apart from a factor $2J+1$. Therefore, 
it is possible to get rid of the central contributions --including the density-dependent terms-- by considering weighted differences as 
\begin{equation}
\frac{1}{2J+1} {\mathcal{V}}(^{2S+1}L_J) - \frac{1}{2J'+1}  {\mathcal{V}}(^{2S+1}L_{J'}) \, .
\end{equation}
In practice, we have considered the following three combinations 
\begin{equation}
\left. 
\begin{array}{l}
\delta_P={\mathcal{V}(^{3}P_1)}/3-{\mathcal{V}(^{3}P_0)}  \\
\delta_D={\mathcal{V}(^{3}D_2)}/5-{\mathcal{V}(^{3}D_3)}/7  \\
\delta_F={\mathcal{V}(^{3}F_3)}/7-{\mathcal{V}(^{3}F_4)}/9  
\end{array}
\right\} 
\label{eq:combination}
\end{equation}
and neglected combinations involving partial waves with $L>3$, which gives smaller contributions. 

We proceed now to give the explicit expressions for these combinations as calculated from several effective interactions. 
The Gogny interaction includes a zero-range spin orbit term 
\begin{eqnarray}
V^G_{LS}({\bf r}) = i W_0 ({\mathbf{\sigma}}_1 + {\mathbf{\sigma}}_2) \cdot 
[{\mathbf k'} \times \delta({\mathbf r}) {\mathbf k} ] \, ,
\label{eq:so}
\end{eqnarray}
where ${\mathbf r}$ is the relative coordinate of the pair, and ${\mathbf k'}$ and ${\mathbf k}$ are the relative momentum operators acting on the left and right, respectively.  
For the tensor interaction we follow the authors of Ref.~\cite{gra13}, which on top of the standard Gogny interaction~\cite{dec80} have added a finite-range tensor effective interaction of the form
\begin{eqnarray}
V^G_{T}(r)=\left[ V_{T1}+V_{T2}P_{\tau}\right] r^2S_T(\hat{\br})\exp^{-r^2/\mu_G^2} \, ,
\label{tensor-gog}
\end{eqnarray}
where $P_{\tau}$ is the isospin exchange operator,
$S_T(\hat{\br})=3(\vsigma_{1} \cdot \hat{\br})(\vsigma_{2} \cdot \hat{\br})-\vsigma_{1}\vsigma_{2}$ is the tensor operator, and a single Gaussian has been used. Due to the chosen isospin structure of the interaction, $V_{T1} + V_{T2}$ is the strength of the force acting between proton-proton or neutron-neutron pairs whereas $V_{T2}$ is the strength of a neutron-proton pair.

The differences (\ref{eq:combination}) we are interested in are written as
\begin{equation}
\left. 
\begin{array}{l}
\delta_P  = -  \frac{W_0}{80} \rho k_F^2  - \frac{27 \mu_G^2 }{40 \sqrt{\pi}} \left( V_{T_1} + V_{T_2} \right)  G_1(k_F \mu_G) \\ 
\delta_D =   - \frac{27 \mu_G^2 }{280 \sqrt{\pi}} \left( V_{T_1} - V_{T_2} \right) G_2(k_F \mu_G) \\  
\delta_F =   - \frac{3 \mu_G^2 }{10 \sqrt{\pi}} \left( V_{T_1} + V_{T_2} \right)  G_3(k_F \mu_G)  
\end{array}
\right\} 
\label{eq:combi-gogny}
\end{equation}
The functions $G_L(k_F \mu_T)$ are given by the sum over momenta on the SNM Fermi sphere of the radial multipole $L$ of the matrix elements of the tensor interaction. They are obtained after some tedious, but straightforward calculations, and are expressed as a polynomial in $a=k_F \mu_G$ plus a combination with polynomial coefficients of the exponential ${\rm e}^{-a^2}$ and the function $S(a)=\gamma - {\rm Ei}(-a^2) + \log (a^2)$, where $\gamma$ is the Euler constant and Ei is the exponential integral function. The first of these functions is written as 
\bea
G_1(a) &=& \frac{1 }{a^3} 
\bigg\{ 6 -6 a^2 - a^4  - 6 {\rm e}^{ - a^2} + 4 a^2 S(a)
\bigg\}  \;.
\eea
The functions with $L>1$ involve higher order polynomials. 

The M3Y interaction~\cite{nak03} is based on the so-called Michigan three-range
Yukawa (M3Y) interaction~\cite{ber97}, which was derived from the bare NN interaction, by fitting the Yukawa functions to
the G-matrix. The central part is completed with a zero-range density dependent term, while the spin-orbit and tensor components are written as 
\be
V^Y_{SO}(r) =  \left[ t^{(LSE)} P_{TE} + t^{(LSO)} P_{TO} \right] \frac{{\rm e}^{-r \mu_{LS}}}{r \mu_{LS}} \, {\bf L}_{12} \cdot {\bf S}_{12} \;,
\ee
\be
V^Y_{T}(r) =  \left[ t^{(TNE)} P_{TE} + t^{(TNO)} P_{TO}
\right] \frac{{\rm e}^{-r \mu_T}}{r \mu_T} \, r^2 S_{T}(\hat{r})  \;,
\label{tensor-naka}
\ee
where $P_{TE}$ and $P_{TO}$ are the triplet-even and triplet-odd operators, and ${\bf L}_{12}={\bf r} \times {\bf p}_{12}$ is the relative orbital angular momentum operator, with ${\bf p_{12}} = i( {\mathbf \nabla}_1 - {\mathbf \nabla}_2)$, and ${\bf S}_{12}$ is the total spin of the pair. Actually, both components contain a superposition of two radial functions with different ranges, and a sum over them is to be understood in the following. 

The differences given in Eq.~(\ref{eq:combination}) are written as
\begin{equation}
\left. 
\begin{array}{l}
\delta_P  =  - \frac{3  t^{(LSO)} }{128 \pi} Y^{(LS)}_1 \left(\frac{k_F}{\mu_{LS}} \right)  -  \frac{27 t^{(TNO)}}{80 \pi \mu_T^2}   Y^{(T)}_1 \left(\frac{k_F}{\mu_{T}} \right)  \\
\delta_D   =   \frac{t^{(LSE)}}{128 \pi}    Y^{(LS)}_2 \left(\frac{k_F}{\mu_{LS}} \right)  - \frac{27 t^{(TNE)}}{560 \pi \mu_T^2}   Y^{(T)}_2 \left(\frac{k_F}{\mu_T} \right) \\
\delta_F   =  \frac{ t^{(LSO)} }{32 \pi}    Y^{(LS)}_3 \left(\frac{k_F}{\mu_{LS}} \right)  - \frac{3 t^{(TNO)}}{40 \pi \mu_T^2}  Y^{(T)}_3 \left(\frac{k_F}{\mu_T} \right) 
\end{array}
\right\} 
\label{eq:combi-nakada}
\end{equation}
where the functions $Y_L^{(LS)}(k_F/\mu_{LS})$ and $Y_L^{(T)}(k_F/\mu_T)$ are given by the sum over momenta on the SNM Fermi sphere of the radial multipole $L$ of the matrix elements of the spin-orbit and tensor interactions, respectively. The first ones are
\bea
Y^{(LS)}_1(a) &=&  \frac{1}{a^3} \bigg\{
-4 a^2 - 88 a^4 +  128 a^3 {\rm ArcTan}(2 a) \nnn
&& \hspace{1cm} + (1 - 24 a^2 + 16 a^4) \log(1 + 4 a^2) \nnn
&& \hspace{1cm} + 16 a^2 {\rm PolyLog}(2, -4 a^2) \bigg\}  \;,
 \\
Y^{(T)}_1(a)  &=& 
\frac{1}{a^3} 
\bigg\{ 4 a^2(3- 2 a^2)  - 3(1+4 a^2 ) \log(1+4 a^2) \nnn
&& \hspace{2cm} - 8 a^2 {\rm PolyLog}(2,-4 a^2)  \bigg\}  \;.
\eea
Similarly to the previous $G_L$ functions, those with $L>1$ involve higher order polynomials in the variables $k_F/\mu_T$ and $k_F/\mu_{LS}$. 

For the Gogny interaction we have used the parametrization D1ST2a~\cite{gra13}, whose single tensor range has been fixed to be equal to the longest range of the Gogny D1S interaction, \emph{i.e.} $\mu_G=1.2$~fm. The spin-orbit and tensor strengths have been obtained by making an adjustment on the neutron $1f$ spin-orbit splitting for the nuclei $^{40}$Ca, $^{48}$Ca, and $^{56}$Ni. The resulting spin-orbit parameters is $W_0=103$ MeV~fm$^5$, a value which is smaller than the original D1ST one ($W_0=130$ MeV~fm$^5$). The tensor parameters are given in Tab.~\ref{tab:D1ST2a}.
For the M3Y interaction we have used the parametrization M3Y-P2, for which the ranges are $\mu_{LS}^{-1}= 0.25$ and 0.4 fm, and $\mu_T^{-1}= 0.4$ and 0.7 fm. The M3Y-P2 set of tensor parameters is fixed so as to reproduce 
the single-particle energy ordering for $^{208}$Pb. Its values are given in Tab.~\ref{tab:M3Y-P2}.

\begin{table}[h]
\begin{center}
\begin{tabular}{c|c|c}
\hline
\hline
 & Ref.~\cite{gra13} & Fitted \\
 \hline
$V_{T1}+V_{T2}$  & -75 & 82.3 \\
$V_{T1}-V_{T2}$  & -195 & -273.9 \\
\hline
\hline
\end{tabular}
\caption{Tensor parameters (in MeV fm$^5$) of interaction D1ST2a~\cite{gra13}. The last column displays the fitted values to the combinations $\delta_P$ and $\delta_D$, as explained in the text.}
\label{tab:D1ST2a}
\end{center}
\end{table}

\begin{table}[h]
\begin{center}
\begin{tabular}{c|c|c}
\hline
\hline
 & Ref.~\cite{nak03} & Fitted \\
 \hline
$t^{TNE}_1$ & -131.52 & 1299.3 \\
$t^{TNE}_2$  & -3.708 & -299.50 \\
$t^{TNO}_1$ & 29.28 & 1734.5 \\
$t^{TNO}_2$  & 1.872 & 29.709 \\
\hline
\hline
\end{tabular}
\caption{Tensor parameters (in MeV fm$^{-2}$) of interaction M3Y-P2~\cite{nak03}.  The last column displays the fitted values to the combinations $\delta_P$ and $\delta_D$, as explained in the text.}
\label{tab:M3Y-P2}
\end{center}
\end{table}

\begin{figure}[h]
\centering
\includegraphics[width=0.47\textwidth]{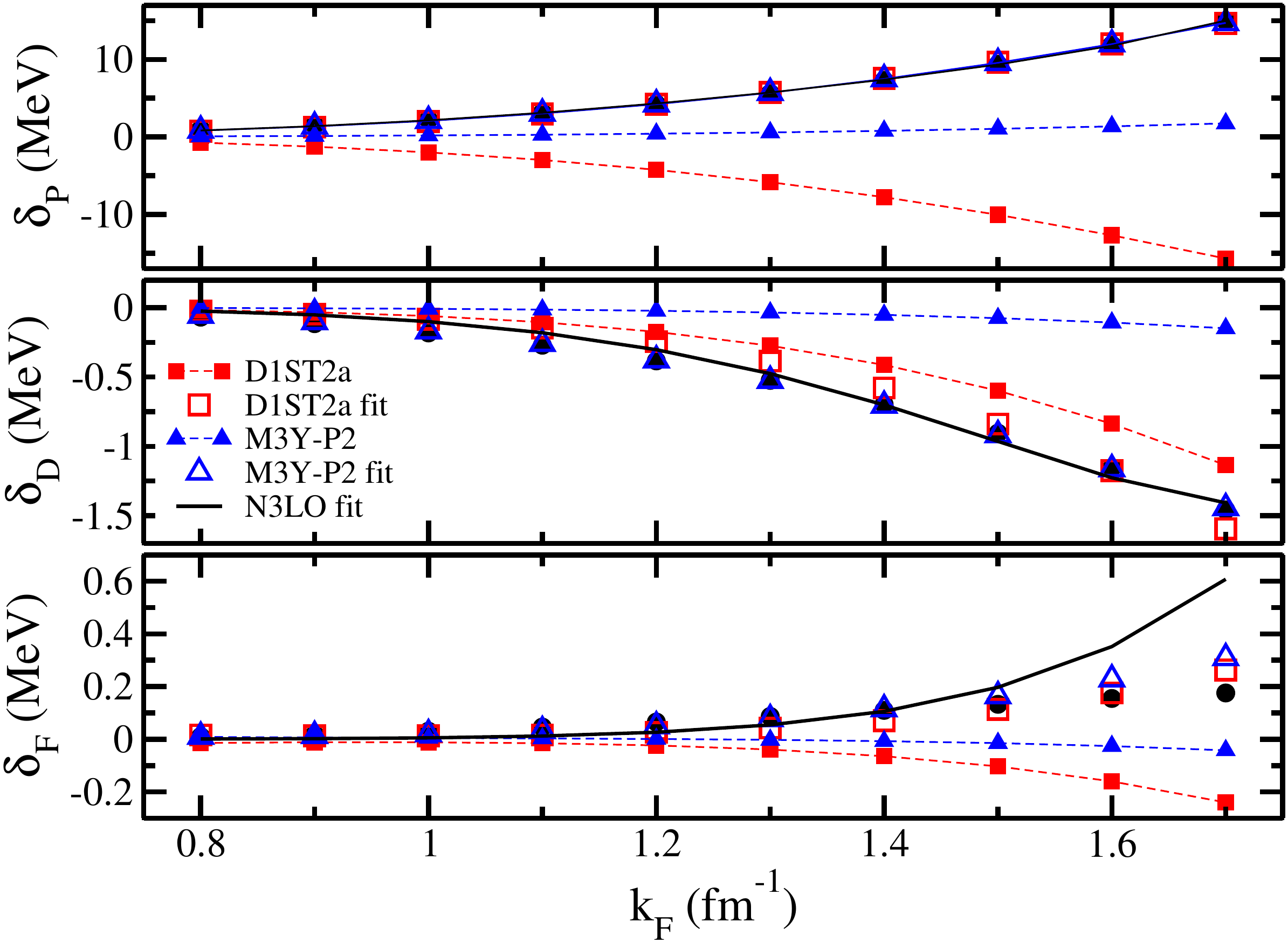}
\caption{(Colors online) Results for the combinations $\delta_P, \delta_D,\delta_F$ defined in Eq.~(\ref{eq:combination}). 
Black circles are the BHF results~\cite{bal97}, while filled red squares and blue triangles have been calculated with interactions D1ST2a~\cite{gra13} and M3Y-P2~\cite{nak03}, respectively. Open symbols and solid line are the result of a fit of the tensor parameters, except for $\delta_F$, as explained in the text.}
\label{JLS-waves}
\end{figure}

In Fig.~\ref{JLS-waves} are displayed the three combinations $\delta_P$, $\delta_D$, and $\delta_F$ as a function of the Fermi momentum $k_F$, with the BHF results~\cite{bal97} represented as black circles. Filled squares and triangles are the results obtained with the interactions D1ST2a and M3Y-P2, respectively. One can see that both interactions lead to results which are in disagreement with BHF ones. For D1ST2a we can roughly say the value of $V_{T1}+V_{T2}$ has the right order of magnitude, but should be changed of sign in order than the calculated $\delta_P$ and $\delta_F$ be in agreement with BHF results. Similarly, the absolute value of $V_{T1}-V_{T2}$ should be slightly increased. As for M3Y-P2 interaction, the comparison shows that while the interaction coefficients produce the right sign for these combinations,  the strengths should be increased in absolute value to get agreement with BHF results. 

Similar drawbacks of D1ST2a and M3Y-P2 interactions have been signaled in Ref. ~\cite{pas14L}, where tensor Landau parameters $H^{(n)}_{\ell}$ for pure neutron matter were calculated with these interactions and compared to the values obtained within the Correlated Basis Function (CBF) of Ref.~\cite{ben13} and the Chiral Effective Field Theory (CEFT) of Ref.~\cite{hol13}.  Although these methods provides results which could differ up to a factor of $\simeq 2$, they all agree on the sign, which is opposite to what has been obtained for the D1ST2a interaction. On the contrary, the results obtained with M3Y-P2 interaction gives values of $H^{(n)}_{\ell}$ which are systematically much smaller than those given by bare-interaction based methods, but with the same sign.  
These points can be quantitatively assessed by fitting the tensor parameters to the BHF results of partial waves. Consider first the corresponding Gogny interaction combinations (\ref{eq:combi-gogny}). As we have realized that the spin-orbit contribution to $\delta_P$ is relatively small, we have kept unchanged $W_0$, and fit the strengths $V_{T1} \pm V_{T2}$ to the BHF $\delta_P$ and $\delta_D$, conserving the same value for the range $\mu_G$. 
The combination $\delta_F$ is not included in the fit, and provides a control about its overall quality. 
The fit has been limited to values of the Fermi momentum relevant for finite nuclei, that is $k_F \le 1.7$~fm$^{-1}$. 
Similarly, for the M3Y-P2 interaction we keep the original spin-orbit parameters as well as the tensor ranges, and have fitted the tensor parameters entering Eqs.~(\ref{eq:combi-nakada}). 
The fitted values of the parameters are given in Tables \ref{tab:D1ST2a} and \ref{tab:M3Y-P2}, and the partial wave results are displayed 
as open symbols with dotted lines in Fig.~\ref{JLS-waves}. 
Comparison of the original and fitted parameters confirms the drawbacks of these effective tensor interactions. In their present form and/or fitting procedure they lead to inconsistencies in the simultaneous description of finite nuclei and infinite matter splittings (see \cite{sta77,gra13}).

Let us now turn to the non-local zero-range pseudo-potential N3LO~\cite{car08,rai11}. Imposing local gauge invariance, the tensor component is written as~\cite{Dav15}
\begin{eqnarray}
V_T^{(N3LO)} & = & \frac{1}{2} t_e T_e +\frac{1}{2} t_o T_o  \nonumber\\
&+&  t_e^{(4)} \left[ (\bk^2+\bkp^2) T_e + 2 (\bk \cdot \bkp) T_o\right]\nonumber\\
&+&  t_o^{(4)} \left[(\bk^2+\bkp^2)T_o + 2 (\bk \cdot \bkp) T_e \right]  \nnn
&+&  t_e^{(6)} \left[ \left( \frac{1}{4} (\bk^2+\bkp^2)^2 +(\bk \cdot \bkp)^2 \right) T_e  \right. \nnn
&& \hskip 1 true cm  \left. + (\bk \cdot \bkp) (\bk^2+\bkp^2)T_o \right]  \nnn
&+& t_o^{(6)} \left[ \left( \frac{1}{4} (\bk^2+\bkp^2)^2 +(\bk \cdot \bkp)^2 \right) T_o  \right.\nnn
&& \hskip 1 true cm  \left. +(\bk \cdot \bkp) (\bk^2+\bkp^2)T_e \right] \, ,
\label{eq:tenseurN3LO}
\end{eqnarray}
where $T_e$ and $T_o$ are the following even and odd tensor operators
\begin{eqnarray}
T_{e}(\mathbf{k'},\mathbf{k})&=&3(\vsigma_{1}\cdot\mathbf{k'})(\vsigma_{2}\cdot\mathbf{k'})+3(\vsigma_{1}\cdot\mathbf{k})(\vsigma_{2}\cdot\mathbf{k})\nnn
&-&(\mathbf{k'}^{2}+\mathbf{k}^{2})(\vsigma_{1}\cdot \vsigma_{2}) \, ,\nnn
T_{o}(\mathbf{k'},\mathbf{k})&=&3(\vsigma_{1}\cdot\mathbf{k'})(\vsigma_{2}\cdot\mathbf{k})+3(\vsigma_{1}\cdot\mathbf{k})(\vsigma_{2}\cdot\mathbf{k}')\nnn
&-&2(\mathbf{k'}\cdot \mathbf{k})(\vsigma_{1} \cdot \vsigma_{2}) \, . \nn
\end{eqnarray}
As for the spin-orbit part, it has been shown~\cite{dav14a} that the form given in Eq.~(\ref{eq:so}) is the only possible one consistent with local gauge invariance. The N3LO combinations (\ref{eq:combination}) then read
\begin{equation}
\left. 
\begin{array}{l}
\delta_P = - \frac{1}{80} k_F^2 W_0 \rho \\
\quad \quad \quad + \frac{9}{160} k_F^2 t_o \rho + \frac{27}{200} k_F^4 t_o^{(4)} \rho + 
 \frac{27}{500} k_F^6 t_o^{(6)} \rho   \\
\delta_D =  \frac{27}{7000} k_F^4 t_e^{(4)} \rho + \frac{9}{3500} k_F^6 t_e^{(6)} \rho  \\
\delta_F =  \frac{1}{875} k_F^6 t_o^{(6)} \rho 
\end{array}
\right\} 
\label{eq:combi-N3LO}
\end{equation}
Presently there is no parametrization of a N3LO pseudo-potential that incorporates finite-nuclei constraints. We have therefore fitted the tensor parameters to the BHF results, by keeping the value for $W_0$ fixed in Ref.~\cite{gra13}. The results are plotted as solid lines in Fig.~\ref{JLS-waves}. It can be seen that a N3LO pseudo-potential contains the relevant degrees of freedom. In fact, the results for combinations $\delta_P$ and $\delta_D$ are of a quality comparable to those obtained from the fits with D1ST2a and M3Y-P2 finite-range expressions. As shown in (\ref{eq:combi-N3LO}), the predicted $\delta_F$ is proportional to $k_F^9$; it is not surprising to get a disagreement with BHF results for values of $k_F \ge 1.5$~fm$^{-1}$.  One should keep in mind that a N3LO pseudo-potential can be viewed as a low-momentum expansion ($k, k' \le 2 k_F$) of a finite-range interaction, as was explicitly shown for a central interaction~\cite{vau72,dob10}.  
Considering the momentum matrix elements of the tensor interactions (\ref{tensor-gog}) and (\ref{tensor-naka}) and
limiting ourselves to the first non-vanishing contribution for $F$-waves, it can be easily shown that in both cases the obtained structure is the same as the N3LO pseudo-potential. Actually, this result is general and it is straightforward to see that Eq.~(\ref{eq:tenseurN3LO}) represents the common low-momentum expansion of any finite-range tensor interaction. 
The N3LO parameters are thus related to those of a finite-range interaction. 

This relation between parameters can be alternatively obtained by 
expanding in powers of $k_F$ the combinations (\ref{eq:combination}) as calculated with a finite-range interaction 
and identifying the coefficients of $k_F$ with the corresponding powers of (\ref{eq:combi-N3LO}). 
Since the spin-orbit term of D1ST2a interaction coincides with the N3LO one, one has simply to Taylor expand the functions $G_L$ entering (\ref{eq:combi-gogny}). One can immediately see that the resulting N3LO coefficients are not independent, since the following equalities hold:  $t_o^{(4)}=- \mu_G^2 t_o$, and $t_{o,e}^{(6)}=- \mu_G^2 t_{o,e}^{(4)}$. That means that the tensor D1ST2a interaction has not the most general form, and at least a second Gaussian term should be included in Eq.~(\ref{tensor-gog}) to get independent N3LO parameters. 
This procedure of relating interaction coefficients is more involved in the case of M3Y-P2 interaction due to its finite-range spin-orbit term. Indeed, in that case the local-gauge invariance is not satisfied, and both spin-orbit and tensor terms contribute to the N3LO partial wave combinations given in Eq.(\ref{eq:combination}). In any case, the N3LO Skyrme pseudo-potential provides a reliable approximation to any finite-range effective interaction in the density range relevant to nuclear structure and astrophysical issues~\cite{dav16}. 

In conclusion, we have shown that N3LO constitutes a reliable substitute of any finite-range tensor interaction and that the tensor parameters of any effective interaction, either finite-range or non-local zero-range, can be obtained from microscopic calculations based on bare NN interactions. In particular, BHF results have been used as an illustrative example, but other methods as $\chi$-EFT or MBPT could be used once these quantities will be made available. These results can be used as a strong constraint, together with finite nuclei results, in the fitting protocole of effective interactions.

\section*{Acknowledgments}

We thank M. Baldo for providing us with the BHF results. We are grateful to J. Meyer and A. Polls for useful discussions. The work of J.N. has been supported by grant  FIS2014-51948-C2-1-P, Mineco (Spain).

\bibliography{biblio}

\end{document}